\documentclass{svproc}
\usepackage{url}

\usepackage{graphicx}
\usepackage{amsfonts}
\usepackage{hyperref}
\usepackage{enumitem}
\usepackage{bbding}
\usepackage{amsmath}

\begin{document}
\mainmatter              
\title{On the Solution of Linear Programming Problems in the Age of Big Data}
\titlerunning{Linear Programming in the Age of Big Data}
\author{{Irina Sokolinskaya \and Leonid B. Sokolinsky\Envelope}\thanks{The
reported study has been partially supported by the RFBR according to research
project \mbox{No.~17-07-00352-a}, by the Government of the Russian Federation
according to Act 211  (contract \mbox{No.~02.A03.21.0011.}) and by the
Ministry of Education and Science of the Russian Federation (government
order 1.9624.2017/7.8).}}
\authorrunning{Irina Sokolinskaya \textit{et al.}} 
%
\tocauthor{Irina Sokolinskaya and Leonid B. Sokolinsky}
\institute{South Ural State University \\76 Lenin prospekt, Chelyabinsk, Russia, 454080\\
\email{Irina.Sokolinskaya@susu.ru}, \email{Leonid.Sokolinsky@susu.ru}}

\maketitle              

\begin{abstract}
The Big Data phenomenon has spawned large-scale linear programming problems. In
many cases, these problems are non\nobreakdash-\hspace{0pt}stationary. In this
paper, we describe a new scalable algorithm called \emph{NSLP} for solving
high-dimensional, non\nobreakdash-\hspace{0pt}stationary linear programming
problems on modern cluster computing systems. The algorithm consists of two
phases: \emph{Quest} and \emph{Targeting}. The \emph{Quest} phase calculates a
solution of the system of inequalities defining the constraint system of the
linear programming problem under the condition of dynamic changes in input
data. To this end, the apparatus of Fejer mappings is used. The
\emph{Targeting} phase forms a special system of points having the shape of an
$n$\nobreakdash-\hspace{0pt}dimensional axisymmetric cross. The cross moves in
the $n$\nobreakdash-\hspace{0pt}dimensional space in such a way that the
solution of the linear programming problem is located all the time in an
$\varepsilon$-vicinity of the central point of the cross.

\keywords{\emph{NSLP} algorithm $\cdot$ non\nobreakdash-\hspace{0pt}stationary linear
programming problem $\cdot$ large-scale linear programming $\cdot$ Fejer mapping.}
\end{abstract}
\section{Introduction}
The Big Data phenomenon has spawned large-scale linear programming (LP)
problems~\cite{sokol_Chung}. Such problems arise in many different fields.
In~\cite{sokol_Tipi}, the following large-scale industrial optimization
problems are presented within the context of big data:
\begin{itemize}
\item schedule crews for 3400 daily flights in 40 countries;

\item buy ads in 10--15 local publications across 40\,000 zip codes;

\item pick one of 742 trillion choices in creating the US National Football
    League schedule;

\item select 5 offers out of 1000 for each of 25\,000\,000 customers of an
    online store;

\item place 1000 stock keeping units on dozens of shelves in 2000 stores;

\item decide among 200\,000\,000 maintenance routing options.
\end{itemize}
Each of these problems uses Big Data from the subject field. Such a problem is
formalized as a linear programming problem involving up to tens of millions of
constraints and up to hundreds of millions of decision variables.

Gondzio~\cite{sokol_Gondzio} presents a certain class of large-scale
optimization problems arising in quantum information science and related to
Bell's theorem. These problems are two-level optimization problems. The
higher-level problem is a non\nobreakdash-\hspace{0pt}convex
non\nobreakdash-\hspace{0pt}linear optimization task. It requires solving
hundreds of linear programming problems, each of which can contain millions of
constraints and millions of variables.

Mathematical modeling in economics is another source of large-scale LP
problems. In many cases, LP problems arising in mathematical economy are
non-stationary (dynamic). For example, Sodhi~\cite{sokol_Sodhi} describes a
dynamic LP task for asset-liability management. This task involves 1.7 billion
constraints and 5.1 billion variables. Algorithmic trading is another area that
generates large-scale non\nobreakdash-\hspace{0pt}stationary linear programming
problems~\cite{sokol_Dyshaev,sokol_Ananchenko,sokol_Radenkov}. In such
problems, the number of variables and inequalities in the constraint system
formed by using Big Data can reach tens and even hundreds of thousands, and the
period of input data change is within the range of hundredths of a second.

Until now, one of the most popular methods for solving LP problems is the class
of algorithms proposed and designed by Dantzig on the basis of the simplex
method~\cite{sokol_Dantzig}. The simplex method has proved to be effective in
solving a large class of LP problems. However, Klee and Minty~\cite{sokol_Klee}
gave an example showing that the worst-case complexity of the simplex method is
exponential time. Nevertheless, Khaciyan~\cite{sokol_Khachiyan} proved that the
LP problem can be solved in polynomial time by a variant of an iterative
ellipsoidal algorithm developed by Shor~\cite{sokol_Shor}. Attempts to apply
the ellipsoidal algorithm in practice have been unsuccessful so far. In most
cases, this algorithm demonstrated much worse efficiency than the simplex
method did. Karmarkar~\cite{sokol_Karmarkar} proposed the \emph{interior-point
method}, which runs in polynomial time and is also very efficient in practice.

The simplex method and the interior-point method remain today the main methods
for solving the LP problem. However, these methods may prove ineffective in the
case of large-scale LP problems with rapidly changing and (or) incomplete input
data. The authors described in~\cite{sokol_Sokolinskaya2010} a parallel
algorithm for solving LP problems with non-formalized constraints. The main
idea of the proposed approach is to combine linear programming and discriminant
analysis methods. Discriminant analysis requires two sets of patterns $M$ and
$N$. The first set must satisfy the non-formalized constraints, while the
second must not. To obtain representative patterns, methods of data
mining~\cite{sokol_Rechkalov} and time series analysis can be
used~\cite{sokol_Zymbler}. To overcome the problem of
non\nobreakdash-\hspace{0pt}stationary input data, the authors proposed
in~\cite{sokol_Sokolinskaya2016_2,sokol_Sokolinskaya2015} the pursuit algorithm
for solving non\nobreakdash-\hspace{0pt}stationary LP problems on cluster
computing systems. The pursuit algorithm uses Fejer mappings
(see~\cite{sokol_Eremin2009}) to build a pseudo-projection onto a convex
bounded set. The pseudo-projection operator is similar to a projection, but in
contrast to the last, it is stable to dynamic changes in input data.
In~\cite{sokol_Sokolinskaya2016}, the authors investigated the efficiency of
using Intel Xeon Phi multi-core processors to calculate the pseudo-projections.

In this paper, we describe the new \emph{NSLP} (Non-Stationary Linear
Programming) algorithm for solving large-scale
non\nobreakdash-\hspace{0pt}stationary LP problems on cluster computing
systems. The \emph{NSLP} algorithm is more efficient than the pursuit
algorithm, since it uses a compute-intensive pseudo-projection operation only
once (the pursuit algorithm computes pseudo-projections $K$ times at each
iteration, $K$ being the number of processor nodes). The rest of the paper is
organized as follows. Section 2 gives a formal statement of an LP problem and
presents the definitions of the Fejer process and the pseudo-projection onto a
polytope. Section 3 describes the new \emph{NSLP} algorithm. Section 4
summarizes the obtained results and proposes directions for future research.

\section{Problem statement} \label{sokol_MathDef}

Let a non\nobreakdash-\hspace{0pt}stationary LP problem be given in the vector
space ${\mathbb{R}^n}$:
\begin{equation}
\label{sokol-Eq1}
\max\left\{ {\left\langle {c_t,x} \right\rangle |{A_t}x \le {b_t},\;x \ge 0} \right\},
\end{equation}
where the matrix ${A_t}$ has $m$ rows. The
non\nobreakdash-\hspace{0pt}stationarity of the problem means that the values
of the elements of the matrix ${A_t}$ and the vectors ${b_t}$, ${c_t}$ depend
on time $t \in {\mathbb{R}_{ \ge 0}}$. We assume that the value of $t=0$
corresponds to the initial time:
\begin{equation} \label{sokol-Eq2}
{A_0} = A, {b_0} = b, {c_0} = c.
\end{equation}

Let us define the map ${\varphi _t}\colon{\mathbb{R}^n} \to {\mathbb{R}^n}$ as
follows:
\begin{equation} \label{sokol-Eq3}
{\varphi _t}\left( x \right) = x - \frac{\lambda }{m}\sum\limits_{i = 1}^m {\frac{{\max \left\{ {\left\langle {{a_{ti}},x} \right\rangle  - {b_{ti}},0} \right\}}}{{{{\left\| {{a_{ti}}} \right\|}^2}}}}  \cdot a_{ti},
\end{equation}
where ${a_{ti}}$ is the $i$-th row of the matrix ${A_t}$, and ${b_{t1}}, \ldots
,{b_{tm}}$ are the elements of the column ${b_t}$. Let us denote
\begin{equation} \label{sokol-Eq4}
\varphi \left( x \right) = {\varphi _0}\left( x \right) = x - \frac{\lambda }{m}\sum\limits_{i = 1}^m {\frac{{\max \left\{ {\left\langle {{a_i},x} \right\rangle  - {b_i},0} \right\}}}{{{{\left\| {{a_i}} \right\|}^2}}}}  \cdot {a_i}.
\end{equation}

Let ${M_t}$ be the polytope defined by the constraints of the
non\nobreakdash-\hspace{0pt}stationary LP problem~(\ref{sokol-Eq1}). Such a
polytope is always convex. It is known (see~\cite{sokol_Eremin2009}) that
${\varphi _t}$ is a continuous single-valued ${M_t}$-fejerian\footnote{A
single-valued map $\varphi\colon{\mathbb{R}^n} \to {\mathbb{R}^n}$ is said to
be \emph{fejerian} relatively to a set $M$ (or briefly, $M$-\emph{fejerian}) if
\[\begin{gathered}
\varphi \left( y \right) = y, \forall y \in M;\hfill\\
\left\| {\varphi (x) - y} \right\| < \left\| {x - y} \right\|, \forall x \notin M,\forall y \in M.\hfill\\
\end{gathered}\]
} map for the relaxation factor $0 < \lambda  < 2$.

By definition, put
\begin{equation} \label{sokol-Eq5}
\varphi _t^s(x) = \underbrace {{\varphi _t} \ldots {\varphi _t}(x)}_s.
\end{equation}

The \emph{Fejer process} generated by the map ${\varphi_ t}$ for an arbitrary
initial approximation ${x_0} \in {\mathbb{R}^n}$ is the sequence $\left\{
{\varphi _t^s({x_0})} \right\}_{s = 0}^{ + \infty }$. It is known (see Lemma
39.1 in~\cite{sokol_Eremin1999}) that the Fejer process for a fixed $t$
converges to a point belonging to the polytope $M_t$:
\begin{equation} \label{sokol-Eq6}
\left\{ {\varphi _t^s({x_0})} \right\}_{s = 0}^{ + \infty } \to \bar x \in {M_t}.
\end{equation}

Let us consider the simplest non\nobreakdash-\hspace{0pt}stationary case, which
is a translation of the polytope $M = {M_0}$ by the fixed vector $d \in
{\mathbb{R}^n}$ in one unit of time. In this case, ${A_t} = A, {c_t} = c$, and
the non\nobreakdash-\hspace{0pt}stationary problem~(\ref{sokol-Eq1}) takes the
form
\begin{equation} \label{sokol-Eq7}
\max\left\{ {\left\langle {c,x} \right\rangle |A(x - td) \le b,\;x \ge 0} \right\},
\end{equation}
which is equivalent to
\[\max\left\{ {\left\langle {c,x} \right\rangle |Ax \le b + Atd,\;x \ge 0} \right\}.\]
Comparing this with~(\ref{sokol-Eq1}), we obtain ${b_t} = b + Atd$. In this
case, the ${M_t}$-fejerian map~(\ref{sokol-Eq3}) is converted to the following:
\[{\varphi _t}\left( x \right) = x - \frac{\lambda }{m}\sum\limits_{i = 1}^m {\frac{{\max \left\{ {\left\langle {{a_i},x} \right\rangle  - \left( {{b_i} + \left\langle {{a_i},td} \right\rangle } \right),0} \right\}}}{{{{\left\| {{a_i}} \right\|}^2}}}}  \cdot {a_i},\]
which is equivalent to
\begin{equation} \label{sokol-Eq8}
{\varphi _t}\left( x \right) = x - \frac{\lambda }{m}\sum\limits_{i = 1}^m {\frac{{\max \left\{ {\left\langle {{a_i},x - td} \right\rangle  - {b_i},0} \right\}}}{{{{\left\| {{a_i}} \right\|}^2}}}}  \cdot {a_i}
\end{equation}

The $\varphi$-\emph{projection} (\emph{pseudo-projection}) of the point $x \in
{\mathbb{R}^n}$ on the polytope $M$ is the map $\pi _M^\varphi (x) = \lim _{s
\to \infty } {\varphi ^s}(x)$.

\section{The \emph{NSLP} algorithm}\label{sokol_Algorithm}

The \emph{NSLP} (\emph{Non-Stationary Linear Programming}) algorithm is
designed to solve large-scale non\nobreakdash-\hspace{0pt}stationary LP
problems on cluster computing systems. It consists of two phases: \emph{Quest}
and \emph{Targeting}. The \emph{Quest} phase calculates a solution of the
system of inequalities defining the constraint system of the linear programming
problem under the condition of dynamic changes in input data. To this end, the
apparatus of Fejer mappings is used. The \emph{Targeting} phase forms a special
system of points having the shape of an $n$\nobreakdash-\hspace{0pt}dimensional
axisymmetric cross. The cross moves in the
$n$\nobreakdash-\hspace{0pt}dimensional space in such a way that the solution
of the LP problem remains permanently in an $\varepsilon$-vicinity of the
central point of the cross. Let us describe both phases of the algorithm in
more detail.

\subsection{The \emph{Quest} phase}

Without loss of generality, we can assume that all the calculations are
performed in the region of positive coordinates. At the beginning, we choose an
arbitrary point ${z_0} \in \mathbb{R}_{ \ge 0}^n$ with
non\nobreakdash-\hspace{0pt}negative coordinates. This point plays the role of
initial approximation for the problem~(\ref{sokol-Eq1}). Then we organize an
iterative Fejer process of the form~(\ref{sokol-Eq6}). During this process, the
Fejer approximations are consecutively calculated by using the Fejer
mapping~(\ref{sokol-Eq3}). This process converges to a point located on the
polytope ${M_t}$. Owing to the non\nobreakdash-\hspace{0pt}stationary nature
of the problem~(\ref{sokol-Eq1}), the polytope ${M_t}$ can change its
position and shape during the calculation of the pseudo-projection. An input
data update is performed every $L$ iterations, $L$ being some fixed positive
integer that is a parameter of the algorithm. Let us denote by ${t_0},{t_1},
\ldots ,{t_k}, \ldots $ sequential time points corresponding to the instants of
input data update. Without loss of generality, we can assume that
\begin{equation} \label{sokol-Eq9}
{t_0} = 0,{t_1} = L,{t_2} = 2L, \ldots ,{t_k} = kL, \ldots.
\end{equation}
This corresponds to the case when one unit of time is equal to the time spent
by the computer to calculate one value of the Fejer mapping using
equation~(\ref{sokol-Eq3}).

\begin{figure}
  \centering
  \includegraphics[scale=0.80]{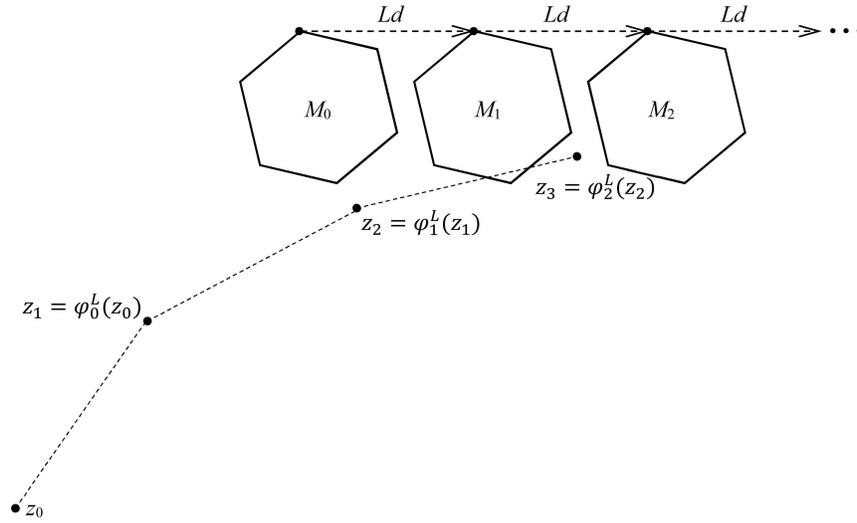}
  \caption{The iterative process in the \emph{Quest} phase
  for problem~(\ref{sokol-Eq7})}
  \label{sokol_Figure1}
\end{figure}

Let the polytope~${M_t}$ take shapes and locations
\[{M_0},{M_1}, \ldots ,{M_k}, \ldots \]
at time points~(\ref{sokol-Eq9}). Let
\[{\varphi _0},{\varphi _1}, \ldots ,{\varphi _k}, \ldots \]
be the Fejer mappings determined by
equation~(\ref{sokol-Eq3}) taking into account the changes in input data of
problem~(\ref{sokol-Eq1}) at time points~(\ref{sokol-Eq9}). In the \emph{Quest}
phase, the iterative process calculates the following sequence of points (see
Fig.\,\ref{sokol_Figure1}):
\[\{ {z_1} = \varphi _0^L({z_0}),{z_2} = \varphi _1^L({z_1}), \ldots ,{z_k} = \varphi _{k - 1}^L({z_{k - 1}}), \ldots \}.\]
Let us briefly denote this iterative process as
\begin{equation} \label{sokol-Eq10}
\left\{ {\varphi _k^L({z_0})} \right\}_{k = 0}^{ + \infty }.
\end{equation}
It terminates when\footnote{Here $\mathrm{dist}(z,M) = \inf
\left\{ {\left\| {z - x} \right\|:x \in M} \right\}$.}
\[\mathrm{dist} \left( {\varphi _k^L({z_{k - 1}}),{M_k}} \right) < \varepsilon,\]
where $\varepsilon  > 0$ is a positive real number being a parameter of the
algorithm. One of the most important issues is the  convergence of the iterative
process~(\ref{sokol-Eq10}). In the general case, this issue remains open.
However, the following theorem holds for the
non\nobreakdash-\hspace{0pt}stationary problem~(\ref{sokol-Eq7}).

\begin{theorem}\label{sokol-th1}
Let a non\nobreakdash-\hspace{0pt}stationary LP problem be given
by~(\ref{sokol-Eq7}). Let the Fejer mappings ${\varphi _0},{\varphi _1}, \ldots
,{\varphi _k}, \ldots$ be defined by the equation
\begin{equation} \label{sokol-Eq11}
{\varphi _k}\left( x \right) = x - \frac{\lambda }{m}\sum\limits_{i = 1}^m {\frac{{\max \left\{ {\left\langle {{a_i},x - kLd} \right\rangle  - {b_i},0} \right\}}}{{{{\left\| {{a_i}} \right\|}^2}}}}  \cdot {a_i}.
\end{equation}
This equation is derived using~(\ref{sokol-Eq8}) and~(\ref{sokol-Eq9}). By definition, put
\begin{equation} \label{sokol-Eq12}
{z_k} = \varphi _{k - 1}^L({z_{k - 1}})
\end{equation}
where $k = 1,2, \ldots $. Then
\begin{equation} \label{sokol-Eq13}
\lim _{k \to \infty } \mathrm{dist}({z_k},{M_k}) = 0
\end{equation}
under the following condition:
\begin{equation} \label{sokol-Eq14}
\forall x \in {\mathbb{R}^n}\backslash M \left( {\left\| {Ld} \right\| < \mathrm{dist} (x,M) - \mathrm{dist} ({\varphi ^L}(x),M)} \right).
\end{equation}
\end{theorem}

The Theorem \ref{sokol-th1} gives a sufficient condition for the convergence of
the iterative process shown in Fig.\,\ref{sokol_Figure1}. To prove this
theorem, we will need the following auxiliary lemma.

\begin{lemma}\label{sokol-le1} Under the conditions of Theorem~\ref{sokol-th1}, we have
\begin{equation} \label{sokol-Eq15}
v - u = pLd \Rightarrow \varphi _p^l(v) - {\varphi ^l}(u) = pLd
\end{equation}
for any $p = 0,1,2, \ldots $, $l = 1,2,3, \ldots $ and $u,v \in {\mathbb{R}^n}$.
\end{lemma}
\begin{figure}
  \centering
  \includegraphics[scale=0.8]{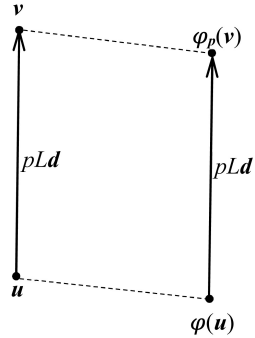}
  \caption{Illustration to the proof of Lemma \ref{sokol-le1}}
  \label{sokol_Figure2}
\end{figure}

\begin{proof}
The proof is by induction on $l$.

Induction base.  Let $l = 1$, then the following condition holds:
\begin{equation} \label{sokol-Eq16}
v - u = pLd.
\end{equation}
Then using~(\ref{sokol-Eq16}), (\ref{sokol-Eq11}) and~(\ref{sokol-Eq4}), we get
\begin{multline*}
{\varphi _p}(v) - \varphi (u) = {\varphi _p}(u + pLd) - \varphi (u) = \\
= u + pLd - \frac{\lambda }{m}\sum\limits_{i = 1}^m {\frac{{\max \left\{ {\left\langle {{a_i},u} \right\rangle  - {b_i},0} \right\}}}{{{{\left\| {{a_i}} \right\|}^2}}}}  \cdot {a_i} - \varphi (u) =\\
= u + pLd - \frac{\lambda }{m}\sum\limits_{i = 1}^m {\frac{{\max \left\{ {\left\langle {{a_i},u} \right\rangle  - {b_i},0} \right\}}}{{{{\left\| {{a_i}} \right\|}^2}}}}  \cdot {a_i} - \\
- u + \frac{\lambda }{m}\sum\limits_{i = 1}^m {\frac{{\max \left\{ {\left\langle {{a_i},u} \right\rangle  - {b_i},0} \right\}}}{{{{\left\| {{a_i}} \right\|}^2}}}}  \cdot {a_i} = pLd.
\end{multline*}
Thus, (\ref{sokol-Eq15}) holds if $l = 1$ (see Fig.\,\ref{sokol_Figure2}).

Inductive step. Assume that condition~(\ref{sokol-Eq16}) is true. Using the
induction hypothesis, we get
\begin{equation}\label{sokol-Eq17}
\varphi _p^{l - 1}(v) - {\varphi ^{l - 1}}(u) = pLd.
\end{equation}
\begin{figure}
  \centering
  \includegraphics[scale=0.8]{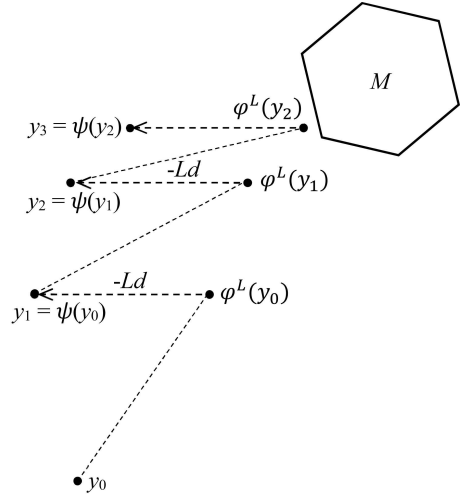}
  \caption{ The process defined by~(\ref{sokol-Eq18})}
  \label{sokol_Figure3}
\end{figure}
Then, combining~(\ref{sokol-Eq5}), (\ref{sokol-Eq17}), (\ref{sokol-Eq11}) and~(\ref{sokol-Eq4}), we obtain
\begin{multline*}
\varphi _p^l(v) - {\varphi ^l}(u) = {\varphi _p}(\varphi _p^{l - 1}(v)) - \varphi ({\varphi ^{l - 1}}(u)) = \\
= {\varphi _p}({\varphi ^{l - 1}}(u) + pLd) - \varphi ({\varphi ^{l - 1}}(u)) = \\
= {\varphi ^{l - 1}}(u) + pLd - \frac{\lambda }{m}\sum\limits_{i = 1}^m {\frac{{\max \left\{ {\left\langle {{a_i},{\varphi ^{l - 1}}(u)} \right\rangle  - {b_i},0} \right\}}}{{{{\left\| {{a_i}} \right\|}^2}}}}  \cdot {a_i} - \varphi ({\varphi ^{l - 1}}(u)) = \\
= {\varphi ^{l - 1}}(u) + pLd - \frac{\lambda }{m}\sum\limits_{i = 1}^m {\frac{{\max \left\{ {\left\langle {{a_i},{\varphi ^{l - 1}}(u)} \right\rangle  - {b_i},0} \right\}}}{{{{\left\| {{a_i}} \right\|}^2}}}}  \cdot {a_i} - \\
- {\varphi ^{l - 1}}(u) + \frac{\lambda }{m}\sum\limits_{i = 1}^m {\frac{{\max \left\{ {\left\langle {{a_i},{\varphi ^{l - 1}}(u)} \right\rangle  - {b_i},0} \right\}}}{{{{\left\| {{a_i}} \right\|}^2}}}}  \cdot {a_i} = pLd.
\end{multline*}
This completes the proof of Lemma \ref{sokol-le1}.
\end{proof}

\begin{proof}[of Theorem \ref{sokol-th1}]
Let us fix an arbitrary point ${z_0} \in {\mathbb{R}^n}\backslash M$. Let the
map $\psi\colon{\mathbb{R}^n} \to {\mathbb{R}^n}$ be given by
\begin{equation}\begin{gathered} \label{sokol-Eq18}
  \psi \left( x \right) = {\varphi ^L}(x) - Ld, \forall x \notin M; \hfill \\
  \psi \left( x \right) = x, \forall x \in M. \hfill \\
\end{gathered}\end{equation}
By definition, put
\begin{equation} \label{sokol-Eq19}
{y_0} = {z_0}
\end{equation}
and
\begin{equation} \label{sokol-Eq20}
{y_k} = \psi ({y_{k - 1}})
\end{equation}
for $k = 1,2, \ldots $ (see Fig.\,\ref{sokol_Figure3}).

\begin{figure}
  \centering
  \includegraphics[scale=0.8]{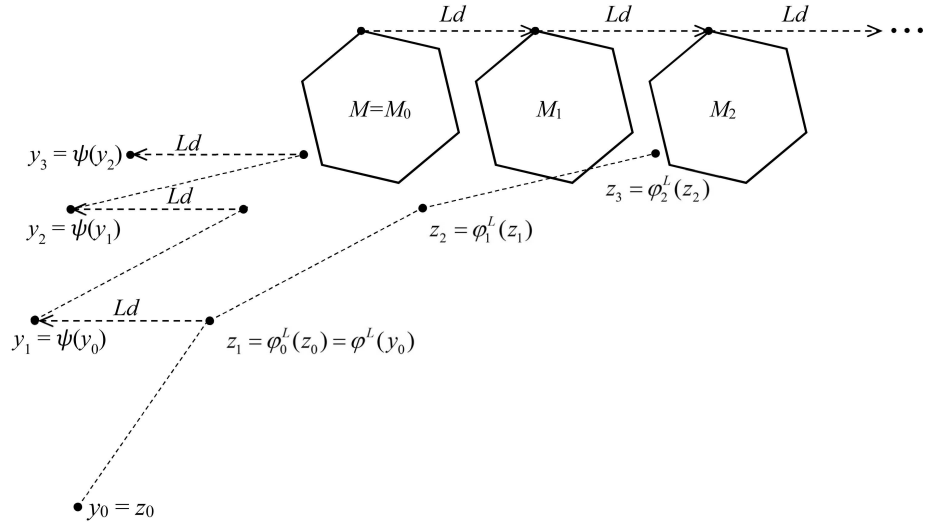}
  \caption{Illustration to equation~(\ref{sokol-Eq21})}
  \label{sokol_Figure4}
\end{figure}

Now let us show by induction on $k$ that
\begin{equation} \label{sokol-Eq21}
{z_k} - {y_k} = kLd
\end{equation}
for $k = 0,1,2,\ldots $ (see Fig.\,\ref{sokol_Figure4}).

Induction base. Equation~(\ref{sokol-Eq21}) holds for $k = 0$. Taking into
account~(\ref{sokol-Eq19}), we see that the equation
\[{z_0} - {y_0} = 0 \cdot Ld\]
holds.

Inductive step. Suppose that
\begin{equation} \label{sokol-Eq22}
{z_{k - 1}} - {y_{k - 1}} = (k - 1)Ld
\end{equation}
for $k > 0$. Substituting $u = {y_{k - 1}}, v = {z_{k - 1}}, l = L, p = k - 1$
in Lemma \ref{sokol-le1}, and using~(\ref{sokol-Eq15}), we obtain
\[{z_{k - 1}} - {y_{k - 1}} = (k - 1)Ld \Rightarrow \varphi _{k - 1}^L({z_{k - 1}}) - {\varphi ^L}({y_{k - 1}}) = (k - 1)Ld.\]
Comparing this with~(\ref{sokol-Eq22}), we have
\begin{equation} \label{sokol-Eq23}
\varphi _{k - 1}^L({z_{k - 1}}) - {\varphi ^L}({y_{k - 1}}) = (k - 1)Ld.
\end{equation}
Combining~(\ref{sokol-Eq20}), (\ref{sokol-Eq18}), (\ref{sokol-Eq12})
and~(\ref{sokol-Eq23}), we get
\begin{multline*}
{z_k} - {y_k} = {z_k} - \psi ({y_{k - 1}}) = {z_k} - {\varphi ^L}({y_{k - 1}}) + Ld ={}\\
{}= \varphi _{k - 1}^L({z_{k - 1}}) - {\varphi ^L}({y_{k - 1}}) + Ld = (k -
1)Ld + Ld = kLd.
\end{multline*}
Thus, equation~(\ref{sokol-Eq21}) holds.

\begin{figure}
  \centering
  \includegraphics[scale=0.75]{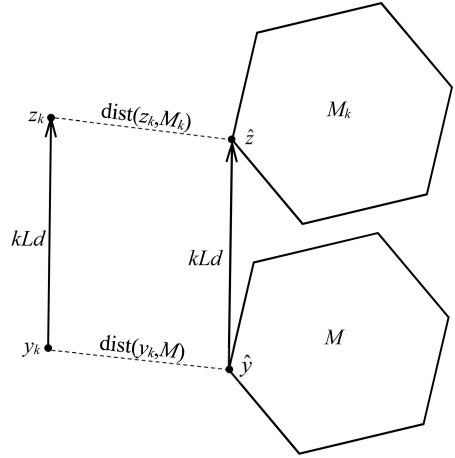}
  \caption{Illustration to equation~(\ref{sokol-Eq24})}
  \label{sokol_Figure5}
\end{figure}

Now we show that
\begin{equation} \label{sokol-Eq24}
\mathrm{dist} ({z_k},{M_k}) = \mathrm{dist} ({y_k},M)
\end{equation}
for all $k = 0,1,2, \ldots $. Let us choose a point $\hat y \in M$ that
satisfies the following condition:
\begin{equation} \label{sokol-Eq25}
\left\| {\hat y - {y_k}} \right\| = \mathrm{dist} ({y_k},M).
\end{equation}
Such a point exists and is unique since the polytope $M$ is a bounded, closed
and convex set. The polytope ${M_k}$ is the result of translating the
polytope $M$ by the vector $kLd$ (see Fig.\,\ref{sokol_Figure5}). Since $\hat
y \in M$, it follows that the point $\hat z = \hat y + kLd$ belongs to the
polytope ${M_k}$. Taking into account~(\ref{sokol-Eq21}), we conclude that
the points $\{ {y_k},{z_k},\hat z,\hat y\} $ are the vertices of a
parallelogram. Therefore,
\begin{equation} \label{sokol-Eq26}
\left\| {\hat z - z_k} \right\| = \left\| {\hat y - {y_k}} \right\|.
\end{equation}
Let us show that
\begin{equation}\label{sokol-Eq27}
\left\| {\hat z - z_k} \right\| =\mathrm{dist}({z_k},{M_k}).
\end{equation}
Assume for a contradiction that $\exists z' \in {M_k}$ such that
\begin{equation} \label{sokol-Eq28}
\left\| {z' - {z_k}} \right\| < \left\| {\hat z - {z_k}} \right\|.
\end{equation}
Since $z' \in {M_k}$, it follows that the point $y' = z' - kLd$ belongs to the
polytope $M$. Now, if we recall that the points $\{ {y_k},{z_k},\hat z,\hat
y\} $ are the vertices of a parallelogram, we get
\[\left\| {y' - {y_k}} \right\| = \left\| {z' - {z_k}} \right\|.\]
Combining this with~(\ref{sokol-Eq28}), (\ref{sokol-Eq26})
and~(\ref{sokol-Eq25}), we obtain
\[\left\| {y' - {y_k}} \right\| = \left\| {z' - {z_k}} \right\| < \left\| {\hat z - {z_k}} \right\| = \left\| {\hat y - {y_k}} \right\| = \mathrm{dist} ({y_k},M).\]
It follows that
\[\exists y' \in M\left( {\left\| {y' - {y_k}} \right\| < \operatorname{dist} ({y_k},M)} \right).\]
This contradicts the definition of the distance between a point and a set.
Therefore, equation~(\ref{sokol-Eq27}) holds. Combining~(\ref{sokol-Eq25}),
(\ref{sokol-Eq26}) and~(\ref{sokol-Eq27}), we get that
equation~(\ref{sokol-Eq24}) also holds.

Further, the map $\psi $ defined by equation~(\ref{sokol-Eq18}) is
single-valued and continuous (this follows from the fact that $\varphi$ is a
single-valued and continuous map). Let us show that the map $\psi $ is
$M$-fejerian. Let $x \in {\mathbb{R}^n}\backslash M$ be an arbitrary point not
belonging to the polytope $M$. Let us choose a point $\hat x \in M$ that
satisfies the following condition
\begin{equation} \label{sokol-Eq29}
\left\| {{\varphi ^L}(x) - \hat x} \right\| = \operatorname{dist} ({\varphi ^L}(x),M).
\end{equation}
Such a point exists and is unique because the polytope $M$ is a bounded,
closed and convex set. Combining the \emph{dist} definition,
equation~(\ref{sokol-Eq18}), the triangle inequality for the norm and
equations~(\ref{sokol-Eq29}) and~(\ref{sokol-Eq14}), we get
\[\begin{gathered}
  \mathrm{dist} (\psi (x),M) \le \left\| {\psi (x) - \hat x} \right\| = \left\| {{\varphi ^L}(x) - Ld - \hat x} \right\| \le  \hfill \\
   \le \left\| {Ld} \right\| + \left\| {{\varphi ^L}(x) - \hat x} \right\| = \left\| {Ld} \right\| + \mathrm{dist} ({\varphi ^L}(x),M) < \mathrm{dist} (x,M). \hfill \\
\end{gathered} \]
It follows that $\psi$ is $M$-fejerian. Therefore,
\[\left\{ {{\psi^k}({y_0})} \right\}_{k = 0}^{ + \infty }\to \bar y \in M.\]
This means that $\mathop {\lim }\limits_{k \to \infty } \operatorname{dist}
({y_k},M) = 0$. Taking into account~(\ref{sokol-Eq24}), we conclude that
$\mathop {\lim }\limits_{k \to \infty } \operatorname{dist} ({z_k},{M_k}) = 0$.
This completes the proof of the theorem.
\end{proof}

From a non-formal point of view, Theorem~\ref{sokol-th1} states that the Fejer
process must converge faster than the polytope $M$ ``runs away''. Manycore
processors can be used to increase the Fejer mapping calculation speed.
In~\cite{sokol_Sokolinskaya2016}, the authors investigated this issue on Intel
Xeon Phi multi-core coprocessors with MIC
architecture~\cite{sokol_Thiagarajan}. It was shown that the Intel Xeon Phi may
be used efficiently for solving large-scale problems.

\subsection{The \emph{Targeting} phase}

The \emph{Targeting} phase begins after the \emph{Quest} phase. At the
\emph{Targeting} phase, an $n$\nobreakdash-\hspace{0pt}dimensional axisymmetric
cross is formed. An $n$\nobreakdash-\hspace{0pt}\emph{dimensional axisymmetric
cross} is a finite set $G = \{ g_0, \ldots,g_P\}  \subset {\mathbb{R}^n}$. The
cardinality of $G$ equals $P + 1$, where $P$ is a multiple of $n \ge2$. The
point ${g_0}$ is singled out from the point set $G$. This point is called the
\emph{central point} of the cross. Initially, the central point is assigned the
coordinates of the point ${z_k}$ calculated in the \emph{Quest} phase by using
the iterative process~(\ref{sokol-Eq10}).

\begin{figure}
\begin{minipage}{0.49\linewidth}
\center{\includegraphics[scale=0.49]{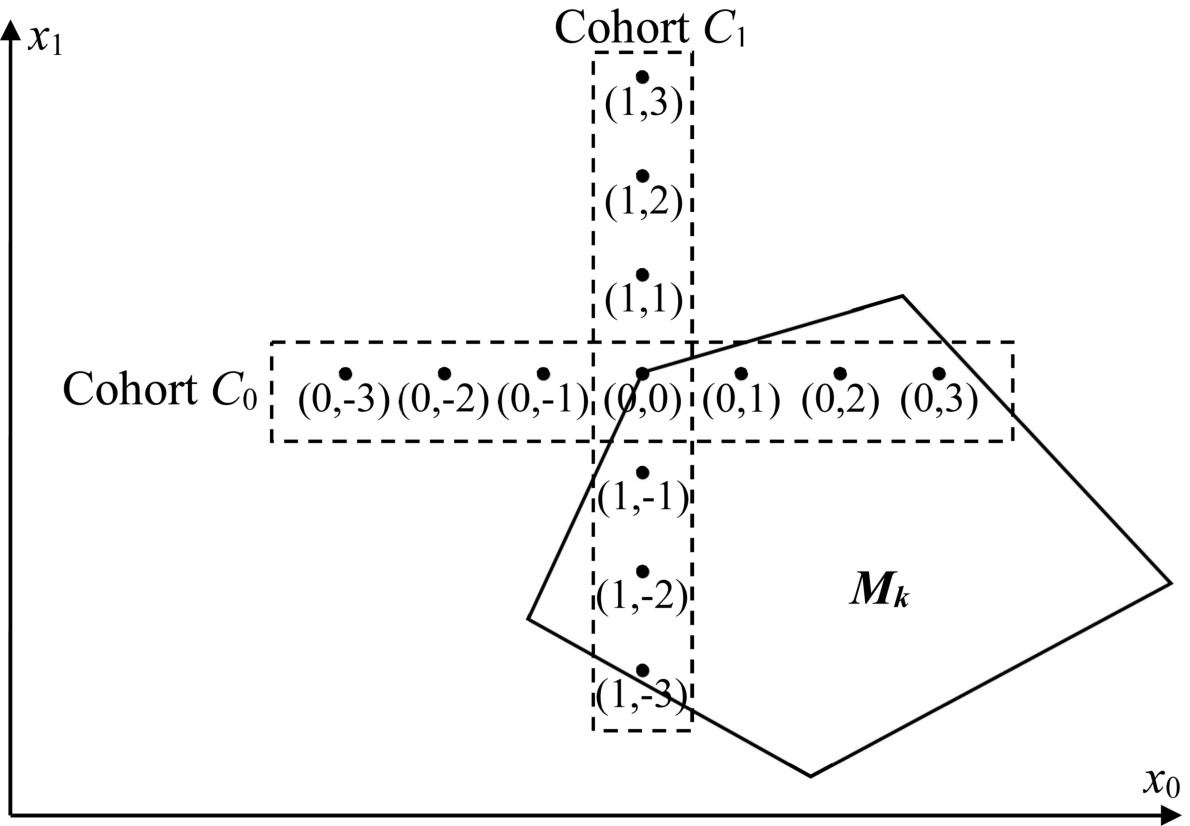}\\ a) with markers $(\chi ,\eta )$}
\end{minipage}
\hfill
\begin{minipage}{0.49\linewidth}
\center{\includegraphics[scale=0.49]{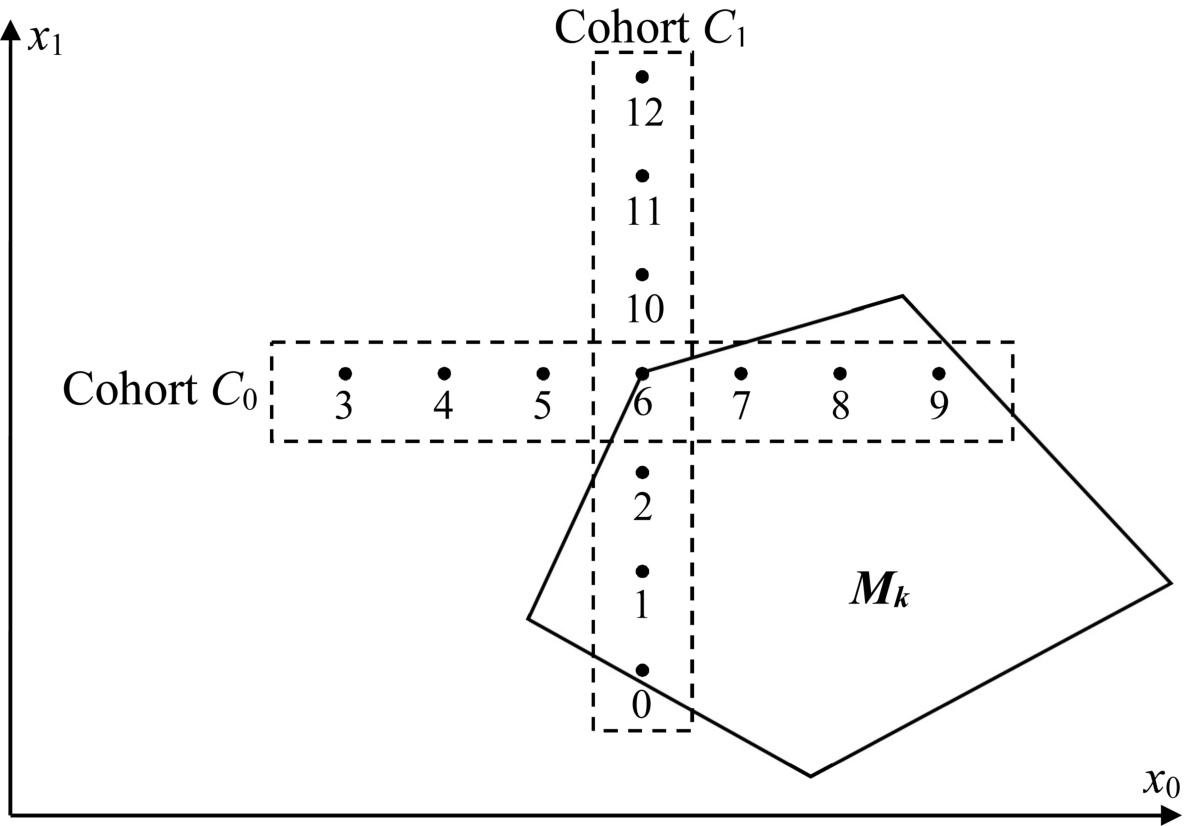}\\ b) sequential numbering}
\end{minipage}
\caption{A two-dimensional cross}
\label{sokol_Figure6-7}
\end{figure}

The set $G\backslash \{ {g_0}\} $ is divided into $n$ disjoint subsets ${C_i}$
($i = 0, \ldots ,n - 1$) called the \emph{cohorts}:
\[G\backslash \{ {g_0}\}  = \bigcup\limits_{i = 0}^{n - 1} {{C_i}} ,\]
where $n$ is the dimension of the space. Each cohort $C_i$ consists of
\begin{equation} \label{sokol-Eq30}
K = \frac{P}{n}
\end{equation}
points lying on the straight line that is parallel to the $i$-th coordinate
axis and passes through the central point ${g_0}$. By itself, the central point
does not belong to any cohort. The distance between any two neighbor points of
the set ${C_i} \cup \{ {g_0}\}$ is equal to a constant~$s$. An example of a
two-dimensional cross is shown in Fig.\,\ref{sokol_Figure6-7}. The number of
points in one dimension, excluding the central point, is equal to $K$. The
symmetry of the cross supposes that $K$ takes only even values greater than or
equal to~2. Using equation~(\ref{sokol-Eq30}), we obtain the following equation
for the total number of points contained in the cross:
\begin{equation} \label{sokol-Eq31}
P + 1 = nK + 1.
\end{equation}
Since $K$ can take only even values greater than or equal to 2 and $n \ge 2$,
it follows from equation~(\ref{sokol-Eq31}) that $P$ can also take only even
values, and $P \ge 4$. In Fig.\,\ref{sokol_Figure6-7}, we have $n = 2$, $K =
6$, $P = 12$.

Each point of the cross $G$ is uniquely identified by a marker being a pair of
integer numbers $(\chi ,\eta )$ such that $0 \le \chi < n$, $\left| \eta
\right| \leq K/2$. Informally, $\chi $ specifies the number of the cohort, and
$\eta $ specifies the sequential number of the point in the cohort ${C_\chi }$
counted from the central point. The corresponding marking of points in the
two-dimensional case is given in Fig.\,\ref{sokol_Figure6-7}~(a). The
coordinates of the point ${x_{(\chi ,\eta )}}$ having the marker $(\chi ,\eta
)$ can be reconstructed as follows:
\begin{equation} \label{sokol-Eq32}
{x_{(\chi ,\eta )}} = {g_0} + (0, \ldots ,0,\underbrace {\eta  \cdot s}_\chi ,0, \ldots ,0).
\end{equation}
The vector being added to ${g_0}$ in the right part of
equation~(\ref{sokol-Eq32}) has a single non-zero coordinate in position $\chi
$. This coordinate equals $\eta  \cdot s$, where $s$ is the distance between
neighbor points in a cohort.

The \emph{Targeting} phase includes the following steps.
\begin{enumerate}
\item Build the $n$\nobreakdash-\hspace{0pt}dimensional axisymmetric cross
    $G$ that has $K$ points in each cohort, the distance between neighbor
    points equaling $s$, and the center at point ${g_0} = {z_k}$, where
    ${z_k}$ is obtained in the \emph{Quest} phase.

\item Calculate $G' = G \cap {M_k}$.

\item Calculate ${C'_\chi } = {C_\chi } \cap G'$ for $\chi  = 0, \ldots ,n
    - 1$.

\item Calculate $Q = \bigcup\limits_{\chi  = 0}^{n - 1} {\left\{ {\arg \max
    \left\{ {\left\langle {{c_k},g} \right\rangle \mid {g \in {{C'}_\chi
    }},{{C'}_\chi } \ne \emptyset } \right\}} \right\}} $.

\item If ${g_0} \in {M_k}$ and $\left\langle {{c_k},{g_0}} \right\rangle
    \ge \mathop {\max }\limits_{q \in Q} \left\langle {{c_k},q}
    \right\rangle $, then $k: = k + 1$, and go to step~2.

\item ${g_0}: = \frac{{\sum\limits_{q \in Q} q }}{{\left| Q \right|}}$.

\item $k: = k + 1$.

\item Go to step~2.
\end{enumerate}

Thus, in the \emph{Targeting} phase, the steps 2--7 form a perpetual loop in
which the approximate solution of the non\nobreakdash-\hspace{0pt}stationary LP
problem is permanently recalculated. From a non-formal point of view, in
Step~2, we determine which points of the cross $G$ belong to the polytope
${M_k}$. To do this, we check the condition ${A_k}g \le {b_k}$ for each point
$g \in G$. Such checks can be executed in parallel by different processor nodes
of a cluster computing system. For this goal to be achieved, $P$
MPI\nobreakdash-\hspace{0pt}processes can be exploited, where $P$ is defined by
equation~(\ref{sokol-Eq31}). We use sequential numbering for distributing the
cross points among the MPI\nobreakdash-\hspace{0pt}processes. Each point of the
cross  is assigned a unique number $\alpha  \in \{ 0, \ldots ,P - 1\} $. The
sequential number $\alpha $ can be converted to a marker $(\chi ,\eta )$ by
means of the following equations\footnote{The symbol $ \div $ denotes integer
division.}:
\[\begin{gathered}
\chi  = \left| {\left| {\alpha  - K} \right| - 1} \right| \div (K/2);\hfill\\
\eta  = \mathrm{sgn} \left( {\alpha  - K} \right) \cdot \left( {\left( {\left( {\left| {\alpha  - K} \right| - 1} \right)\bmod \left( {K/2} \right)} \right) + 1} \right).\hfill\\
\end{gathered}\]
The backward conversion can be performed by means of the equation
\[\alpha  = \eta  + \mathrm{sgn} (\eta )\frac{\chi }{2}K + K.\]
Fig.\,\ref{sokol_Figure6-7}~(b) demonstrates the sequential numbering of points
that corresponds to the marking in Fig.\,\ref{sokol_Figure6-7}~(a).

\section{Conclusion}

In this paper, a new \emph{NSLP} algorithm for solving
non\nobreakdash-\hspace{0pt}stationary linear programming problems of
large dimension has been described. This algorithm is oriented to cluster
computing systems with manycore processors. The algorithm consists of two
phases: \emph{Quest} and \emph{Targeting}. The \emph{Quest} phase calculates a
solution of the system of inequalities defining the constraint system of the
linear programming problem under the condition of input data dynamic changes.
To do this, we organize a Fejer process that computes a pseudo-projection onto
the polytope $M$ defined by the constraints of the LP problem. In this case,
input data changes occur during calculation of the pseudo-projection. A
convergence theorem for the described iterative process is proved in the case
of translation of the polytope $M$. The \emph{Targeting} phase forms a
special system of points having the shape of an
$n$\nobreakdash-\hspace{0pt}dimensional axisymmetric cross. The cross moves in
the $n$\nobreakdash-\hspace{0pt}dimensional space in such a way that the
solution of the linear programming problem is located all the time in an
$\varepsilon$-vicinity of the central point of the cross. A~formal description
of the \emph{Targeting} phase is presented in the form of a sequence of steps.
Our future goal is a parallel implementation of the \emph{NSLP} algorithm in
the C++ language using the MPI library, as well as the development of
computational experiments on a cluster computing system using synthetic and
real LP problems.

\end{document}